# A New Method for Compensation and Rematch of Cavity Failure in the C-ADS Linac [*]

XUE Zhou(薛舟) [1;2] DAI JianPing(戴建枰) [2;1)] MENG Cai(孟才) [2]
1 (University of Chinese Academy of Sciences, Beijing 100049, China)
2 (Institute of High Energy Physics (IHEP), CAS, Beijing 100049, China)

**Abstract**： For proton linear accelerators used in applications such as accelerator-driven systems, due to the nature of the operation, it is essential for the beam failure rate to be several orders of magnitude lower than usual performance of similar accelerators. A fault-tolerant mechanism should be mandatorily imposed in order to maintain short recovery time, high uptime and extremely low frequency of beam loss. This paper proposes an innovative and challenging way for compensation and rematch of cavity failure using fast electronic devices and Field Programmable Gate Arrays (FPGAs) instead of embedded computers to complete the computation of beam dynamics. A method of building an equivalent model for the FPGA, with optimization using a genetic algorithm, is shown. Results based on the model and algorithm are compared with TRACEWIN simulation to show the precision and correctness of the mechanism.

## 1. Introduction

The China Accelerator Driven subcritical System (C-ADS) aims to build a superconducting linac with beam energy of 1.5 GeV and beam current of 10 mA [1]. As a high power proton accelerator, the C-ADS linac should have extremely high availability and reliability all the time [2, 3], as shown in Table 1. This is because unexpected beam trips may lead to serious change of temperature and thermal stress in the reactor core and result in permanent damage of the facilities.

Table 1: C-ADS Design Parameters

| Parameters | Design Value | |
|---|---|---|
| Particle | proton | |
| Energy | 1.5 | GeV |
| Current | 10 | mA |
| Beam Power | 15 | MW |
| RF Frequency | (162.5)/325/650 | MHz |
| Duty Factor | 100 | % |
| Beam Loss | <1 | % |
| Beam Trips/Year | <25000 | 1s<t≤10 |
| | <2500 | 10s<t≤5min |
| | <25 | t>5min |

Reliability-oriented design practices need to be followed from the early design stage. In particular [4]: (1) a high degree of redundancy needs to be planned for critical areas, using methods such as "hot-stand-by" [3]. (2) "strong design" is needed. (3) fault-tolerance capabilities have to be considered, which requires some main components to allow compensation and rematch. The representative compensation work may be found on SNS [5], which uses the global compensation. When the cavities fail, the machine will look up the database to find the compensation data and then readjust the parameters of the working cavities. In C-ADS, some modeling programs to build the database step by step have been used and tested by simulation software for beam dynamics, like TRACEWIN or TRACK [6] [7] [8]. During the calculations of compensation and rematch for each component, a lot of work needs to be prepared by humans, and it is easy to make mistakes during data processing. This paper gives an alternative method, with calculation done in Field Programmable Gate Arrays (FPGAs) to deal with cavity failure. Using FPGAs will decrease the number of repeated calculations and possibility of making mistakes, and shorten the time for compensation and rematch. Due to the limitations of hardware like FPGAs, a high-level algorithm and special model which only includes simple arithmetical operations and logical operations should be built. This paper focuses on the above-mentioned model and algorithm.

[*] Supported by China ADS Project (XDA03020600) and Natural Science Foundation of China (11575216)
1) E-mail: jpdai@ihep.ac.cn

## 2. A new method for compensation and rematch of RF cavity failures

To avoid the beam loss caused by superconducting RF cavity failures, it is necessary to re-adjust the accelerating fields and synchronous phases of the non-faulty cavities to recover the beam. The usual way to implement this is: when the cavities fail, the machine looks up a database which was built by TRACEWIN or other simulation tools in advance and then re-adjusts the parameters of the working cavities [4][9]. Looking up a huge database wastes a lot of time even compared with readjusting hardware, especially through EPICS [10]. Moreover, much more work needs to be done when the lattice changes during operation and the database does not include such situation.

In this paper, we propose an innovative way of calculating online instead of looking up a database to achieve the compensation by a hardware implementation of the scheme using fast electronic devices and FPGAs. This mechanism has its advantages compared with the traditional method, detailed as follows.

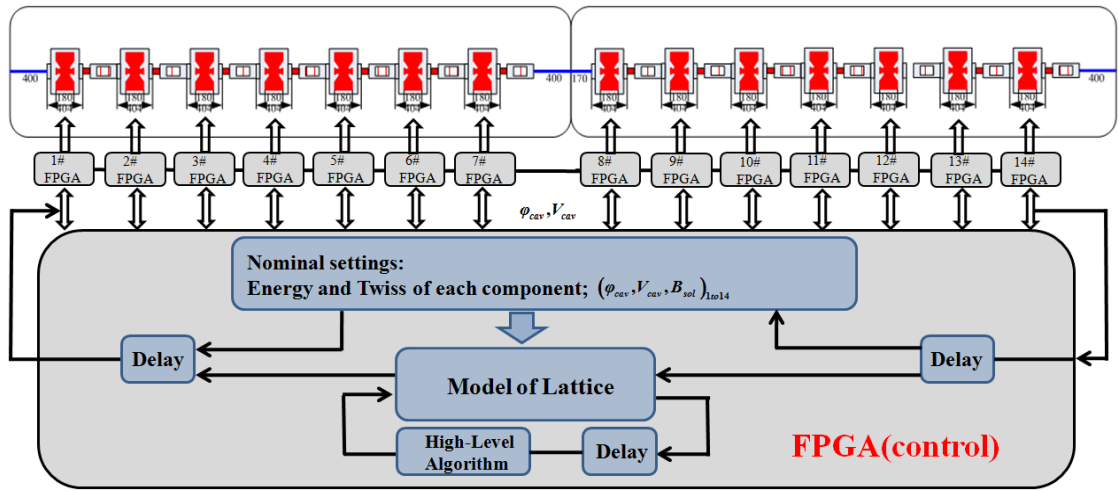

Figure 1: Diagram of the hardware compensation and rematch in Injector I.

(1) Arithmetic computing speed is higher. As an integrated circuit device consisting of logic gates, an FPGA is able to realize parallel calculating and synchronous processing, which means a group of solutions will be found in each clock period after the pipe-line is full. The time to find a global optimum-solution of compensation and rematch will be reduced greatly, which means calculating online is viable compared with the traditional method.

(2) Instantaneous compensation and rematch is easier. Not only is the computing speed higher for FPGAs, but it is also an easier way to connect with the low level RF system, experimental physics and industrial control system, and other types of hardware facilities on the accelerator to make instantaneous compensation and rematch possible.

(3) Good portability and repeatability. Calculating by FPGAs can operate independently of some specific components. When the lattice changes, all the data in the database need to be re-calculated, which means it is necessary to prepare sets of the database. However, it is much easier to get new results with the FPGA, because changing the lattice for the FPGA just means replacing modules. The new method has advantages for subsequent modification and upgrade.

In order to verify this new method, we chose Injector I of C-ADS at the Institute of High Energy Physics (IHEP), Beijing as a test bench. This machine is a 10 MeV proton linac containing fourteen superconducting RF cavities and solenoids. A diagram of the operating principle is shown in Fig.1.

Modeling the lattice is the first step to get a global optimal solution of compensation and rematch. Based on the nominal parameters, continuous iteration of the model can then be carried out by the FPGAs. The optimum solutions are transported as digital signals to re-adjust the elements, while the nominal setting is transported to the lattice during normal operation.

## 3. Equivalent model of dynamic simulation

Due to the limitations of hardware circuits and logic gates, it is hard for FPGAs to realize some complicated operations, such as division, exponent and square root, as opposed to addition and multiplication. Therefore, we choose linear basis function models [11] to mitigate this problem, as shown in Eq. 1.

$$y(x,w) = w_0 + \sum_{j=1}^{M-1} w_j \varphi_j(x) \qquad (1)$$

where $y(x,w)$ is the equivalent model of each transfer matrix element. $\phi_j(x)$ are known as basis functions, which may be replaced by nonlinear functions. $w_j$ are the weights of the basis functions. Each component has its own transfer matrix except cavities and solenoids with actual field, so a structure of "drift + gap/solenoid + drift" is chosen to approximately replace them. Because of this, we can use a polynomial to take place of the transfer matrix of each component. The equivalent model described in Eq.1 was first implemented in MATLAB [12], and then applied to the FPGAs. The simulation waveform in Chip-Scope describes the beam characteristics spreading in the lattice of Injector I, as shown in Fig. 2. When the clock is under 200 MHz, the calculation of longitudinal twiss parameters and for the whole lattice energy takes 270 ns, and horizontal twiss parameters takes 695 ns.

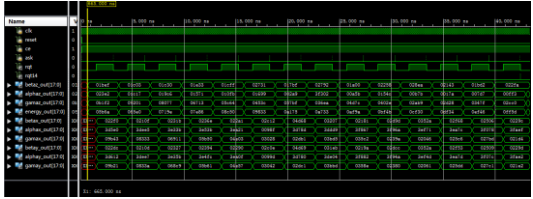

Figure 2: Timing simulation of Injector I

Space charge is an important and complicated factor which brings about coupling between the longitudinal and horizontal beam characteristics. Considering the linear space charge, the equivalent model should divide each component into short slices for which space charge can be dealt with as a thin lens. Similarly, the effect of space charge can be modeled as transfer matrices inserted into each component [13] [14] [15] [16], which means the polynomial model is a suitable format for this scenario. We take the longitudinal result of the equivalent model as an example, shown in Fig.3. The envelopes calculated by the polynomial model and TRACEWIN are nearly coincident.

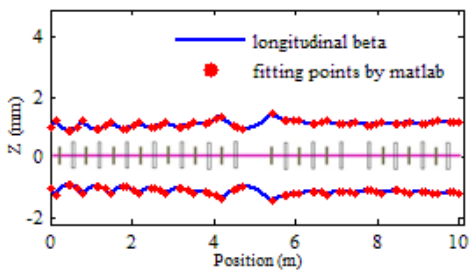

Figure 3: The longitudinal envelopes in TRACEWIN (blue line) and β calculated using polynomial model (red points).

The relative errors between the result from the model and TRACEWIN are shown in Fig.4.

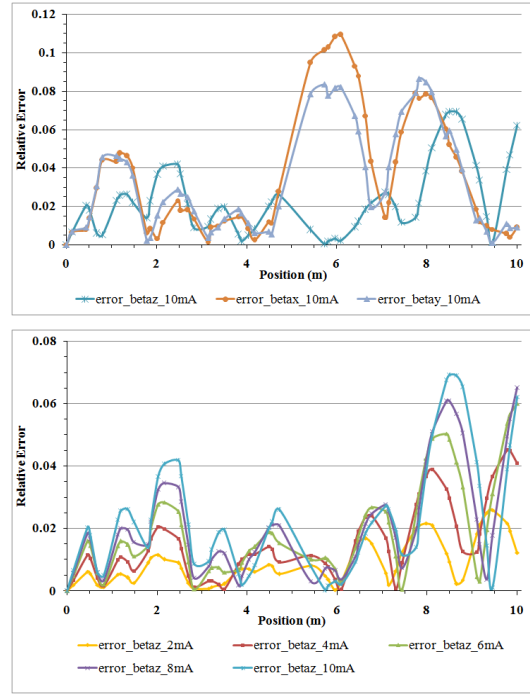

Figure 4: Relative error of beta at 10mA and longitudinal beta at different beam current.

With the beam current increasing, the relative error also shows an obvious increase caused by two factors. First, the effect of space charge shows more strongly in low-energy sections, which makes the linear space charge model insufficient for describing the effect. Secondly, in order to be eventually calculated on FPGAs, the model has to be simplified, which sacrifices the model precision and limits the active zone. However, the distance of local compensation and rematch is less than 4 meters including five periods, which means the relative error can be controlled within five percent during the optimization for Injector I.

## 4. Compensation and rematch algorithm

Cavity failures cause not only loss of energy but also mismatching which eventually leads to beam loss. Re-adjusting neighboring cavities and solenoids may avoid this situation. How to re-adjust becomes a difficult problem, which can be solved by combining the equivalent model with some algorithms. Genetic algorithms [17] can get near-optimal solutions by iteration. A flowchart for a genetic algorithm applied to an FPGA is shown in Fig. 5.

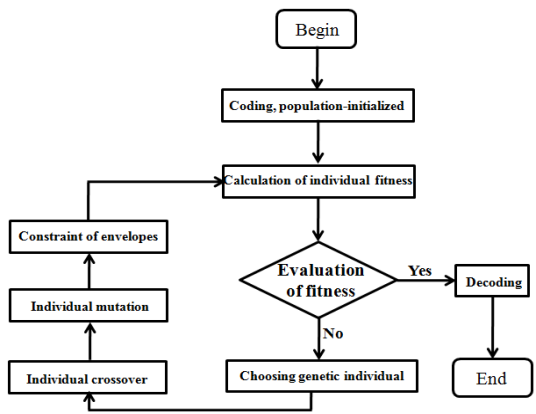

Figure 5: Flowchart for a genetic algorithm.

Within the scope of the active zone of the equivalent model, we randomly produce accelerating fields and synchronous phases of cavities which attend the compensation and rematch. Repeating this whole process N times is known as the initialization of population. Under the same condition of kinetic energy and twiss parameters at the entrance, we can get the state of the beam at matching point N times. As opposed to nominal energy, phase and twiss parameters at the matching point, the best result of this iteration is selected and reserved. At the same time, the limiting condition in the model of longitudinal and horizontal envelopes eliminates the worst solutions during the selection. A typical generation algorithm operator called the *roulette wheel* is used to select high-probability individuals. Subsequently, *single-point intersection* and *mutation* are also applied in the algorithm to generate new individuals. This algorithm chooses the square root of quadratic sum of relative errors to be the object function. Reaching the condition of the specified fitness level or the maximum number of generation will terminate the algorithm.

## 5. TRACEWIN verification of the model and algorithm

In order to verify the feasibility of the hardware compensation and rematch, we used TRACEWIN to test the optimization result using the above-mentioned model and algorithm. It is difficult to realize the compensation and rematch in the low-energy section of the linac [4].

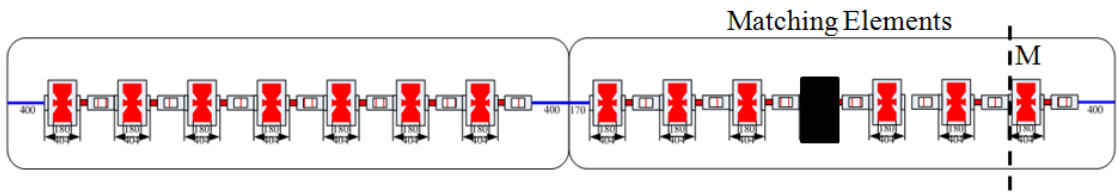

Figure 6: The compensation and rematch for the eleventh cavity's failure.

Therefore, we take a cavity failure in the eleventh period at which point the energy has already reached about 8 MeV, as an example and use the cavities from the ninth period to the thirteenth period to complete the compensation and rematch, as shown in Fig 6. Combining the model and genetic algorithm, we can get the settings of cavities and solenoids for compensation and rematch. We then put these settings in TRACEWIN and obtain the energy and emittance before and after re-adjustment, as shown in Fig. 7.

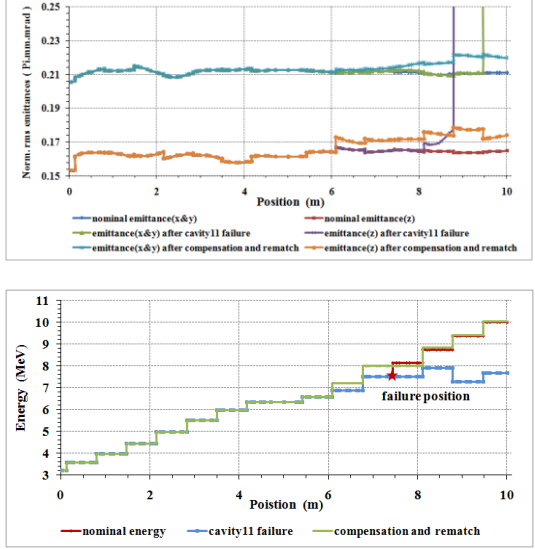

Figure 7: Energy and emittance before and after compensation and rematch.

When nothing is done after cavity 11 fails, the longitudinal emittance shoots up to 10 π.mm.mrad which goes off the scale in Fig.7. After the compensation and rematch, beam energy has recovered to nominal energy (10 MeV). Meanwhile, longitudinal and horizontal emittances show about 5.6% and 4.2% increases respectively at the end of Injector I. Figure 8 shows the envelopes before and after re-adjustment.

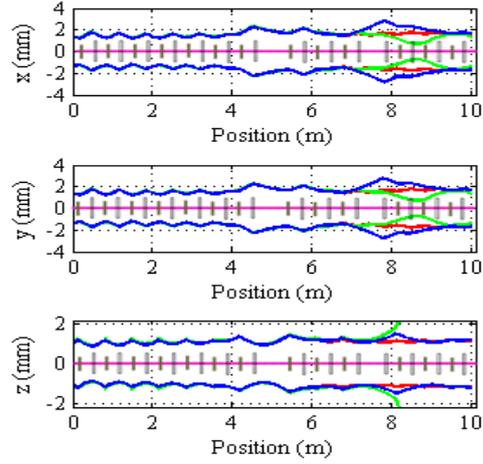

Figure 8: Envelopes before and after compensation and rematch. The red lines are nominal envelopes and the green lines are the envelopes after cavity 11 failure. The blue lines are the envelopes after compensation and rematch.

The longitudinal envelopes after cavity failure have gone off the scale due to the rapid growth. The blue lines are the envelopes after compensation and rematch. The envelopes after compensation and rematch show slight growth near the failing cavity, and later on after the thirteenth period the envelopes are just the same as the nominal lattice. The detailed results for the twiss parameters at the matching point are shown in Table 2. A mismatched beam with mismatch factor [18] under 10% is tolerable especially at low energy.

**Table 2: Twiss parameters at the matching point.**

| Twiss parameters | nominal | After compensation and rematch | Mismatch Factor |
|:---:|:---:|:---:|:---:|
| Beta-x | 1.9548 | 2.0718 | 3.07% |
| Alpha-x | 0.5476 | 0.5974 | |
| Beta-y | 1.9856 | 2.1156 | 3.23% |
| Alpha-y | 0.5599 | 0.5911 | |
| Beta-z | 1.2822 | 1.2408 | 6.53% |
| Alpha-z | -0.3446 | -0.4531 | |

As shown above, the method of equivalent model and genetic algorithm to deal with compensation and rematch efficiently is feasible and effective.

## 6. Conclusion

A new method for compensation and rematch with an equivalent model and high-level algorithm in FPGAs is proposed and has been verified as viable. Using failure of the eleventh cavity in Injector I as an example, the polynomial model with space charge and optimized genetic algorithm has been tested against results calculated by combining Matlab and TRACEWIN with reasonably good results.

## 7. References

[1] Zhihui Li, et al, Physics design of an accelerator for an accelerator-driven subcritical system, Physical Review Special Topics- Accelerators and Beams 16, 080101 (2013).

[2] Tang Jing-Yu and Li Zhi-Hui et al. Edited, Conceptual Physics design on the C-ADS accelerator, IHEP-CADS-Report/2012-01E.

[3] Z.H. Li et al., "BEAM DYNAMICS OF CHINA ADS LINAC", proceeding of HB2012, Beijing, China, THO3A02, http://jacow.org/.

[4] Jean-Luc Biarrotte and Didier Uriot, "Dynamic compensation of an rf cavity failure in a superconducting linac", Physical review special topics-Accelerators and beams, 2008, 11(072803):1-11.

[5] J. Galambos, S. Henderson, Y. Zhang, A Fault Recovery System For the SNS Superconducting Cavity LINAC, Proceedings of LINAC , Knoxville, Tennessee USA  2006.

[6] SUN Biao, YAN Fang et al. Compensation-rematch for the major components of C-ADS injector-I, Chinese Physics C, vol.39 (11), (2015)

[7] D. Uriot, TraceWin Documentation, CEA/SACLAY-DSM/Irfu/SACM, 2014.

[8] K. R. Crandall and D. P. Rusthoi, TRACE 3-D Documentation, LA-UR-97-886, Third Edition, May 1997.

[9] J. Galambos, S. Henderson, A. Shishlo, Y. Zhang, "Operational Experience of a Superconducting Cavity Fault Recovery System at the Spallation Neutron Source," Proceedings of the Workshop on Utilisation and Reliability of High Power Proton Accelerators (HPPA5), 6-9 May 2007, Mol, Belgium, p. 161 (2008).

[10] EPICS, http://www.aps.anl.gov/epics/

[11] M. Jordan, J. Kleinberg, B. Schölkopf. Pattern Recognition and Machine Learning [M]. Springer Science + Business Media, LLC, 2006:138-139.

[12] Jie Chen, Matlab Manual, Publishing House of Electronics Industry, 2013: 152-158.

陈杰, MATLAB 宝典, 电子工业出版社, 2013: 152~158.

[13] Frank J. Sacherer. Rms Envelope Equations with Space Charge, IEEE Trans: 1105-1107 (1971).

[14] LI Jin-Hai, TANG Jing-Yu, OUYANG Hua-Fu. Matching by solenoids in space charge dominated LEBTs. Chinese Physics C, vol.33, No.10:905-913, 2009.

[15] G. Poplau, U. van Rienen. An efficient 3D Space Charge Routine with Self-Adaptive Discretization. Proceedings of ICAP09, San Francisco,

CA.

[16] P. Sing Babu, A. Goswami, V.S. Pandit. Optimization of beam line parameters for space charge dominated multi-species beam using random search method. Physics Letters A 376 (2012) 3192–3198.

[17] David E. Goldberg, Genetic Algorithms in search, Optimization, and Machine Learning, Addison-Wesley, 1989

[18] J. Guyard and M. Weiss, Use Of Beam Emittance Measurements In Matching Problems, Proc. 1976 Proton Linear Accel. Conf., Chalk River, Ontario, Canada.